\documentclass[journal]{IEEEtran}

\usepackage[export]{adjustbox}
\usepackage[dvipsnames]{xcolor}
\usepackage{empheq}
\usepackage{amsmath,amsfonts,amsthm,amssymb,amsbsy,paralist,ifthen}
\usepackage{mathrsfs}
\usepackage[mathcal]{euscript}
\usepackage{graphicx}
\usepackage{psfrag}
\usepackage{subfigure}
\usepackage{url}
\usepackage{stfloats}
\usepackage{float}
\usepackage{array}
\usepackage{booktabs} 
\usepackage{algorithm} 
\usepackage{algorithmic} 
\usepackage{color}
\usepackage{cases}
\usepackage{bm}
\usepackage{colortbl}
\usepackage{arydshln}
\usepackage{multirow}
\usepackage{multicol}
\usepackage{cite}
\usepackage{extarrows}
\usepackage[table]{xcolor}

\usepackage{hyperref}

\newtheorem{Lem}{Lemma}

\newtheorem{The}{Theorem}

\newtheorem{Rek}{Remark}

\begin{document}


\title{Spatially Adaptive SWIPT with Pinching Antenna under Probabilistic LoS Blockage}

\author{Ruihong Jiang, Yanqing Xu, Huimin Hu, and Yang Lu

\vspace{-5.5mm}

\thanks{This work is supported by the National Natural Science Foundation of China (NSFC) under Grant No. 62301077.}

\thanks{R. Jiang is with the State Key Laboratory of Networking and Switching Technology, Beijing University of Posts and Telecommunications, Beijing 100876, China (e-mail: rhjiang@bupt.edu.cn).}

\thanks{Y. Xu is with the School of Science and Engineering, The Chinese University of Hong Kong, Shenzhen, 518172, China (email: xuyanqing@cuhk.edu.cn).}

\thanks{H. Hu is with the School of Communications and Information Engineering, Xi’an University of Posts and Telecommunications, Xi’an, 710121, China. (e-mail: huiminhu@xupt.edu.cn).}

\thanks{Yang Lu is with the State Key Laboratory of Advanced
Rail Autonomous Operation, and also with the School of Computer Science and Technology, Beijing Jiaotong University, Beijing 100044, China (e-mail: yanglu@bjtu.edu.cn).}

}

 \maketitle

\begin{abstract}
This paper considers a power-splitting (PS)-based simultaneous wireless information and power transfer (SWIPT) system employing a reconfigurable pinching antenna (PA) under probabilistic line-of-sight (LoS) blockage. We formulate a joint optimization of the PA position and PS ratio to maximize the average signal-to-noise ratio (SNR) at the user, subject to its average energy harvesting (EH) and PA placement range. We derive the closed-form solution. Results show that the EH requirement has a deterministic impact on the optimal PA position and its feasible region, requiring the PA close to the user for large channel gain. Moreover, stronger waveguide attenuation lowers the overall SNR and shifts the optimal PA toward the feed point, while heavier LoS blockage degrades the SNR uniformly with little change in the optimal PA position. Spatial PA adaptation combined with dynamic PS ensures robust SWIPT performance, and mechanical reconfigurability enhances sustainability by guaranteeing energy feasibility in dynamic environments.

\end{abstract}


\section{Introduction}

The rapid evolution of wireless networks has led to an increasing demand for energy‑efficient and self‑sustainable communication systems. Simultaneous wireless information and power transfer (SWIPT) has been considered as a promising paradigm, enabling devices to harvest energy from radio‑frequency signals while decoding information simultaneously \cite{eh_jrh}. However, practical SWIPT systems face significant challenges in dynamic environments, particularly under time‑varying blockage conditions that degrade both communication reliability and energy harvesting (EH) efficiency \cite{nearfieldswipt}.

To enhance SWIPT, flexible antenna architectures have been widely explored. Among them, pinching‑antenna (PA) systems have emerged as a promising solution owing to their ability to mechanically activate conductive pinchers along dielectric waveguides, thereby flexibly adjusting resonant modes and radiation directions \cite{liu_1, Ding_1, lu_1}. Compared with conventional multiple-input multiple-output (MIMO) or reconfigurable intelligent surface (RIS) architectures, PAs can directly create or strengthen line-of-sight (LoS) links on demand, achieving highly adaptable coverage and efficient utilization of spatial degrees of freedom \cite{ding2025flexible,xu2025rate,liu_2, lu_2}. Such properties make PAs an attractive candidate for integrating SWIPT into next‑generation sustainable networks.


Recent studies have explored the benefits of PAs in wireless powered communication networks (WPCN) and SWIPT systems \cite{Li2025PASS_SWIPT, Li2025PA_WPCN, Papanikolaou2025PASS_WPCN, Li2025PA_SWIPT}. Specifically, joint PA positioning and beamforming were optimized in PA-aided MIMO SWIPT \cite{Li2025PASS_SWIPT}, while resource allocation and PA placement were studied in PA-aided WPCNs \cite{Li2025PA_WPCN, Papanikolaou2025PASS_WPCN}. Additionally, PA-assisted SWIPT frameworks demonstrated performance gains via spatial and power domain co-design \cite{Li2025PA_SWIPT}. However, most existing works mainly focus on SWIPT systems with separated information decoding (ID) and EH receivers or time-switching based WPCNs, assuming deterministic channels. Motivated by this, our work study a power splitting (PS)-based SWIPT system under stochastic LoS blockage, addressing the trade-off between EH and ID that is absent in separated architectures.


Despite the advances in PA-aided SWIPT, the assumption of a persistent LoS link is increasingly unrealistic for high-frequency systems (e.g., mmWave or THz), where links are susceptible to stochastic blockage from moving obstacles and environmental dynamics. More recently, the studies \cite{Ding2025LoS} and \cite{xu2025blockage} addressed this issue by establishing a theoretical foundation that analyzes the impact of LoS blockage and waveguide attenuation on the outage probability and the average data rate of the system. Specifically, these works indicate that PAs tend to move closer to users to mitigate path loss when waveguide attenuation is negligible. Yet, these analyses only consider information-related metrics, which are insufficient for SWIPT, as PA placement is governed by dual constraints of ID and EH. Therefore, how to exploit PA reconfigurability for SWIPT under random LoS blockage remains a critical problem, requiring a delicate balance between the conflicting objectives of ID and EH.

To fill the gaps, we propose a novel PA-assisted SWIPT framework designed for dynamic environments with probabilistic LoS blockage. Unlike existing studies focusing on deterministic channels or single metrics, we introduce a stochastic optimization framework that jointly designs the PA position and PS ratio to maximize the average signal-to-noise ratio (SNR) while ensuring minimum average harvested power. Our contributions are summarized as follows:

\vspace{-0.5mm}
\begin{itemize}
    \item We present a PA-assisted SWIPT system that exploits the mechanical reconfigurability of PA to enhance both ID and EH under dynamic blockage conditions. 
    
    \item We formulate an optimization problem to maximize the average SNR by jointly designing the PA position and PS ratio subject to average EH and PA placement limits, capturing the fundamental trade-off between EH and ID.
          
    \item We derive a closed-form solution to the non-convex problem, which is proven to achieve the optimal performance, providing valuable design insights.
\end{itemize}
\vspace{-0.5mm}

Numerical results validate the theoretical solution and show that the PA-enabled system significantly outperforms conventional fixed-antenna deployments in terms of average system performance. The performance gain is benefited by spatial reconfigurability to adapt to blockage statistics and energy constraints, demonstrating a new dimension of resilience for sustainable SWIPT.

\section{System Model}

\begin{figure}[t!]
    \centering    
    \includegraphics[width=0.75\columnwidth]{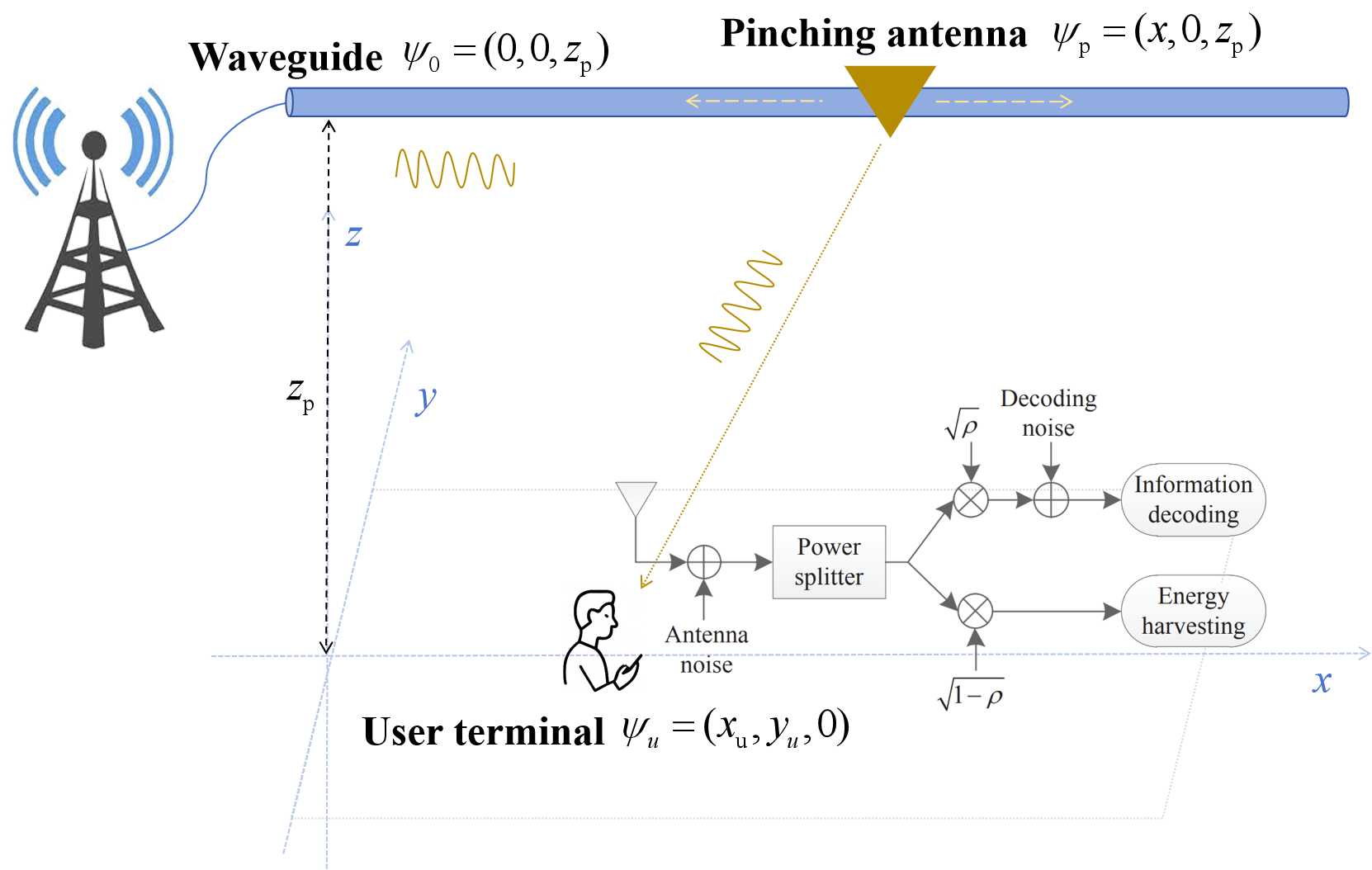}
    \caption{A depicted illustration of the considered system, where a PA serves a downlink SWIPT user in the presence of LoS blockage.}
    \label{fig_sys}
\end{figure}

\subsection{System Network}
Consider a dielectric waveguide of length $L$, excited by a fixed transmitter (feed point) located at $\boldsymbol{\psi}_0 = (0, 0, z_p)$, as shown in Fig. \ref{fig_sys}. The waveguide serves as the primary medium for guiding electromagnetic waves from the transmitter to the desired location. A movable PA slides along the waveguide and can be positioned at any point $x \in [0, L]$ with physical coordinates $\boldsymbol{\psi}_{\rm p} = (x, 0, z_p)$.

The PA acts as a controllable radiating element that couples energy from the waveguide’s guided mode into free space, enabling SWIPT to the user. By adjusting the PA’s position along the waveguide, the system achieves spatial reconfigurability to adapt to dynamic channel conditions and LoS blockages. The transmitted signal is then directed towards the user located at $\boldsymbol{\psi}_{\rm u} = (x_{\rm u}, y_u, 0)$, where $x_{\rm u}$ is aligned with the waveguide axis. At the user, a PS-based SWIPT architecture is employed with a PS ratio $\rho \in [0,1]$. Specifically, a fraction $\rho$ of the received power is allocated to ID, and the remaining fraction $(1-\rho)$ is used for EH. Such a flexible transmission paradigm facilitates the user to simultaneously decode information and harvest energy from the same received signal.

\subsection{Channel Model with Probabilistic LoS Blockage}

The effective channel in the system is influenced by two main components: the feed-to-PA propagation along the waveguide and the PA-to-user radiation in free space.
When the LoS path is unobstructed, the channel coefficient between the PA and the user is given by\footnote{Waveguide attenuation is neglected for tractability, as existing results show its impact on the achievable rate is minor \cite{xu2025pinching, xu2025blockage}.}
\begin{equation}
    h^{\text{LoS}} = \tfrac{\eta^{\frac{1}{2}} e^{-j \frac{2 \pi}{\lambda} ||\boldsymbol{\psi}_{\rm u} - \boldsymbol{\psi}_{\rm p} || + \frac{2 \pi}{\lambda_g} ||\boldsymbol{\psi}_{\rm 0} - \boldsymbol{\psi}_{\rm p}||} }{ || \boldsymbol{\psi}_{\rm u} - \boldsymbol{\psi}_{\rm p} || },
\end{equation}
where $\eta = \frac{c^2}{(4\pi f_c)^2}$ with $c$ being the speed of light and $f_c$ being the carrier frequency. $\lambda$ and $\lambda_g$ denote the free-space and waveguide wavelengths, respectively.
Consequently, the squared complex channel gain follows the free-space path loss low, given by $|h^{\text{LoS}}|^2 = \eta/d^2(x)$,
where $d(x) = \sqrt{(x_{\rm u} - x)^2 + y_{\rm u}^2 + z_{p\rm }^2}$.



In practical wireless environments, the LoS link is subject to random blockage, which is commonly caused by obstacles and user mobility, particularly in high-frequency communication systems such as mmWave and THz. To capture this uncertainty, we adopt a probabilistic LoS blockage model. Specifically, the LoS condition is modeled as a Bernoulli random variable $\gamma \in \{0,1\}$, where $\gamma=1$ represents the presence of a direct LoS path and $\gamma =0$ indicates blockage. The probability that the LOS link exists is expressed as
\begin{flalign}
    \Pr(\gamma = 1) = e^{- \beta d^2(x)}, \label{los_r}
\end{flalign}
where $\beta \in (0, 1]$ is the blockage density parameter that characterizes the environmental clutter. A larger $\beta$ implies a denser obstacle environment, thus reducing the likelihood of maintaining an unobstructed LoS path. Accordingly, the instantaneous channel coefficient for the PA-assisted SWIPT system under the probabilistic LoS blockage can be given by
\begin{flalign}
    h = \gamma h^{\rm LoS}.
\end{flalign}
where the non-line-of-sight (NLoS) contribution is neglected, since NLoS propagation in millimeter-wave and higher frequency bands typically suffers from severe attenuation, making its impact negligible in future high-frequency communication systems \cite{Ding2025LoS, xu2025blockage}.



\subsection{Average Performance Metrics under Stochastic Channels}


Let $P_{\rm t}$ denote the transmit power. The power of the instantaneous received signal for ID and EH can be given by
\begin{align}
P_{\rm r, ID} =& \rho P_{\rm t} |h|^2 = \rho P_{\rm t} |\gamma h^{\rm LoS}|^2, \\
P_{\rm r, EH} =& (1-\rho) P_{\rm t} |h|^2 = (1-\rho)P_{\rm t} |\gamma h^{\rm LoS}|^2, 
\end{align}
respectively. To analyse the performance, we evaluate the expectation of the received ID and EH by averaging over the LoS blockage indicator $\gamma$. That is,
\begin{align}
    f(x) \triangleq \mathbb{E}[|h|^2]  = & \,\, \mathbb{E}_\gamma\left[|\gamma h^{\text{LoS}}|^2\right]  \label{fa} = |h^{\text{LoS}}|^2 \cdot \mathbb{E}_{\gamma}[\gamma^2]  \\
   \overset{(\ref{fa}a)}{=} & \,\,  |h^{\text{LoS}}|^2 \cdot \Pr(\gamma = 1)  
    \overset{(\ref{fa}b)}{=} \,\, \tfrac{\eta e^{-\beta d^2(x)}}{d^2(x)}, \nonumber
\end{align}
where $(\ref{fa}a)$ is due to $\mathbb{E}[\gamma] =  \mathbb{E}[\gamma^2]$ for $\gamma \in \{0,1\}$, and $(\ref{fa}b)$ substitutes coordinates and the LoS probability defined in \eqref{los_r}.

Therefore, the average SNR can be given by
\begin{flalign}
    \overline{\text{SNR}} = \mathbb{E}_\gamma\left[\tfrac{P_{\rm r, ID}}{\sigma^2}\right] = \Gamma \rho \mathbb{E}_\gamma\left[|\gamma h^{\text{LoS}}|^2\right] = \Gamma \rho f(x),
   \label{snr_ave}
\end{flalign}
where $\Gamma = P_{\rm t} / \sigma^2$ with $\sigma^2$ being the noise power at the user. Correspondingly, the average harvested power is given by
\begin{equation}
    \overline{Q}_{\text{EH}} = \mathbb{E}_\gamma\left[ \zeta P_{\rm r, EH} \right] = \zeta(1 - \rho) P_{\rm t} f(x),
\end{equation}
where $\zeta \in (0,1]$ is the energy conversion efficiency.

\section{Problem Formulation and Solution}
We aim to maximize the average SNR\footnote{Average SNR enables closed-form analysis and approximates rate-optimal design, offering a conservative bound for outage performance under blockage.} subject to the minimum average EH requirement $q_0$ and physical constraints on the PA position $x$ and PS ratio $\rho$. The joint optimization problem is formulated as
\begin{subequations}
\begin{align}
\textbf{P}_0: \quad \max_{x,\, \rho} \quad & \Gamma \rho f(x) \\
\text{s.t.} \quad & \zeta (1 - \rho) P_{\rm t} f(x) \ge q_0, \label{eq:eh_const} \\
& 0 \le \rho \le 1, \\
& 0 \le x \le L.
\end{align}
\end{subequations}

Due to the coupling between the PA position $x$ and the PS ratio $\rho$, as well as the probabilistic LoS blockage embedded in $f(x)$, the problem $\mathbf{P}_0$ is non-convex. Nevertheless, its structure, specifically the monotonic dependence of the average harvested power on $\rho$, allows the joint optimization to be decoupled. For any given $x$, the average EH constraint uniquely determines the maximum feasible $\rho$, leaving the average SNR as a function of $x$ alone. This reduces the problem to a one-dimensional optimization over the PA position $x$, as formalized in the following lemma.

\begin{Lem} \label{lem1}
The problem $\mathbf{P}_0$ is feasible if and only if there exists $x \in [0, L]$ such that $f(x) \ge q_0 / (\zeta P_{\rm t})$. Under the feasibility condition, the optimal PS ratio for a given $x$ is
\begin{equation}
    \rho^\star(x) = 1 - \tfrac{q_0}{\zeta P_{\rm t} f(x)}, \label{eq:rho_opt_x}
\end{equation}
and maximizing the average SNR is equivalent to maximizing $f(x)$ over the feasible set of $x$.
\end{Lem}

\begin{proof}
From the average EH constraint \eqref{eq:eh_const}, we have
\begin{flalign}
\zeta(1 - \rho) P_{\rm t} f(x) \ge q_0 \quad \Rightarrow \quad \rho \le 1 - \tfrac{q_0}{\zeta P_{\rm t} f(x)}. 
\end{flalign}
Since the average SNR in \eqref{snr_ave}  is strictly increasing in $\rho$, the optimal $\rho^\star(x)$ achieves the upper bound, i.e., \eqref{eq:rho_opt_x}. Feasibility then requires $\rho^\star(x) \ge 0$, which yields $f(x) \ge \frac{q_0}{\zeta P_{\rm t}}$.

Substituting $\rho^\star(x)$ into the objective function gives
\[
\overline{\text{SNR}} = \Gamma f(x) \left(1 - \tfrac{q_0}{\zeta P_{\rm t} f(x)}\right) = \Gamma \left( f(x) - \tfrac{q_0}{\zeta P_{\rm t}} \right),
\]
which is a strictly increasing function of $f(x)$. Therefore, maximizing $\overline{\text{SNR}}$ is equivalent to maximizing $f(x)$ over all $x \in [0, L]$ satisfying $f(x) \ge \frac{q_0} {\zeta P_{\rm t}}$. The proof is finished.
\end{proof}


\subsubsection{The case without waveguide attenuation}
Based on Lemma \ref{lem1}, the optimal solutions to problem $\textbf{P}_0$ in the absence of waveguide attenuation are derived in Theorem \ref{the1}.

\begin{The} \label{the1}
For the SWIPT-enabled PA system under probabilistic LoS blockage and without waveguide attenuation, the optimal PA position $x^\star$ and PS ratio $\rho^\star$ can be given by
\begin{equation}
\left\{
\begin{aligned}
x^\star &= \operatorname{clip}\left(x_{\rm u}, \max\{0, x_{\rm u} - R\}, \min\{L, x_{\rm u} + R\}\right), \\
\rho^\star &= 1 - q_0/(\zeta P_{\rm t} f(x^\star)),
\end{aligned}
\right. 
\label{the_eq1}
\end{equation}
where $\operatorname{clip}(a, l, u) \triangleq \{l, \, \text{if } a<l; \, a, \, \text{if } a \in [l, u]; \, u, \, \text{if } a>u\}$ projects the argument $a$ onto the feasible interval $[l, u]$, $R = \sqrt{t_{\rm{th}} - y_{\rm u}^2 - z_{\rm p}^2}$ with $t_{\text{th}} = \frac{1}{\beta} W\left(\frac{\beta \eta \zeta P_{\rm t}}{q_0}\right)$ and $W(\cdot)$ being the Lambert $W$ function.
\end{The}


\begin{proof}
Define $t \triangleq (x-x_{\rm u})^2 + y_{\textrm u}^2 + z_{\textrm p}^2>0$. The average channel gain can be written as $f(x) = \eta \tfrac{e^{-\beta t}}{t}$. Let $g(t)=\tfrac{e^{-\beta t}}{t}$.
Taking its first derivative yields
\begin{flalign}
g'(t)=-\tfrac{e^{-\beta t}(\beta t+1)}{t^2}<0, \quad \forall \, t>0,
\end{flalign}
which implies that $f(x)$ is strictly decreasing in $t$. Thus, maximizing $f(x)$ is equivalent to minimizing $t$, i.e., placing the PA as close as the user as possible in the unconstrained case. The EH constraint imposes a strict upper limit on $t$. Substituting $f(x)$ into the feasibility condition, we obtain
  $\eta \tfrac{e^{-\beta t}}{t} \ge \tfrac{q_0}{\zeta P_{\textrm t}}$.  
Then, multiplying both sides by $\beta$ and rearranging terms yields
\begin{flalign}
   (\beta t)e^{\beta t} \le \tfrac{\beta \eta \zeta P_{\textrm t}}{q_0}.
\end{flalign}
By the definition of the Lambert $W$ function, implies $\beta t \le W\left(\frac{\beta \eta \zeta P_{\textrm t}}{q_0}\right)$. Thus, the feasible region for $t$ is given
\begin{flalign}
t \le \tfrac{1}{\beta}W\!\big(\tfrac{\beta\eta\zeta P_{\rm t}}{q_0}\big). \label{t_0}
\end{flalign}
Let $t_{\mathrm{th}} = \frac{1}{\beta}W\!\left(\frac{\beta\eta\zeta P_{\rm t}}{q_0}\right)$ and substituting $t$ into \eqref{t_0}, we have
\begin{flalign}
(x - x_{\rm u})^2 \le t_{\rm{th}} - (y_{\textrm u}^2 + z_{\textrm p}^2). 
\end{flalign}
Because the left-hand side is non-negative, a feasible $x$ exists only if $t_{\text{th}} \ge y_{\rm u}^2 + z_{\rm p}^2$, which is the feasibility condition of problem $\mathbf{P}_0$.
Under this condition, we define $R = \sqrt{t_{\rm{th}} - y_{\rm u}^2 - z_{\rm p}^2}$ and the feasible interval of $x$ becomes
\begin{flalign}
\mathcal{X}_{\mathrm{feasible}} = [x_{\rm u}-R,\,x_{\rm u}+R] \cap [0,L].
\end{flalign}

Since $f(x)$ decreases with $t$ monotonically, the optimal solution is the point in $\mathcal{X}_{\text{feasible}}$ closest to $x_{\rm u}$, which is exactly given by the clip function \eqref{the_eq1}. The optimal PS ratio $\rho^\star$ follows directly from Lemma \ref{lem1}. The proof ends.
\end{proof}



\begin{Rek} \label{rem1}
Without waveguide attenuation, the optimal PA position $x^\star$ reflects a trade-off between maximizing the channel gain and satisfying the EH constraint. Specifically, the parameters $\beta$ and $q_0$ impose a strict feasible region that restricts how closely the PA can approach the user, thereby balancing communication performance and energy sustainability. The optimal PS ratio $\rho^\star$ is uniquely determined by the tight EH constraint. Theorem \ref{the1} provides a baseline for the SWIPT-enabled PA system under probabilistic LoS blockage.
\end{Rek}


\subsubsection{The case with waveguide attenuation}

By introducing the exponential attenuation term $e^{2\alpha x}$, the average channel gain is given by
\begin{flalign}
   f(x) = \tfrac{\eta e^{-\beta\left[(x-x_{\rm u})^2 + A\right]}}{\left[(x-x_{\rm u})^2 + A\right] e^{2\alpha x}}, 
\end{flalign}
where $A = y_{\rm u}^2 + z_{\rm p}^2$ and $\alpha>0$ denotes the waveguide attenuation coefficient.

\begin{The} \label{the2}
For the case with waveguide attenuation, the optimal PA position $x^\star$ is given by
\begin{flalign}
    x^\star = x_{\rm u} + z^*,
\end{flalign}
where $z^*$ is the unique real root of the cubic equation
\begin{flalign}
\beta z^3 + \alpha z^2 + (\beta A + 1)z + \alpha A = 0, \quad z = x - x_{\rm u}. \label{in_2}
\end{flalign}
It holds that $x^\star \le x_{\rm u}$, i.e., $x^\star$ shifts toward the feed point.
\end{The}

\begin{proof}
Maximizing $f(x)$ is equivalent to maximizing $\ln f(x)$. Taking the first-order derivative and setting it to zero yields
\begin{flalign}
    (x-x_{\rm u})\!\left(\beta + \tfrac{1}{(x-x_{\rm u})^2 + A}\right) = -\alpha.
\end{flalign}
Since $\alpha$, $\beta$ and $A$ are non-negative, we have $(x - x_\mathrm{u}) \le 0$, i.e., $x^\star \le x_{\rm u}$. This means $x^\star$ shifts toward the feed point, which agrees with the results in \cite{w_1}. Substituting $z = x - x_{\rm u}$ results in the cubic equation \eqref{in_2}. The closed-form expression for $z^*$ can be obtained via Cardano's method. The derivation is omitted due to page limits.
\end{proof}

\begin{Rek}
When $\alpha=0$ (without waveguide attenuation), $x^\star$ approaches $x_{\rm u}$ to minimize free-space path loss. For $\alpha > 0$, $x^\star$ shifts toward the feed point. This indicates a trade-off where the system sacrifices marginal free-space path gain to reduce waveguide attenuation, thus maximizing total received power.
\end{Rek}

\begin{figure*}[t!]
  \centering
  \subfigure[]{
      \includegraphics[width=0.23\textwidth]{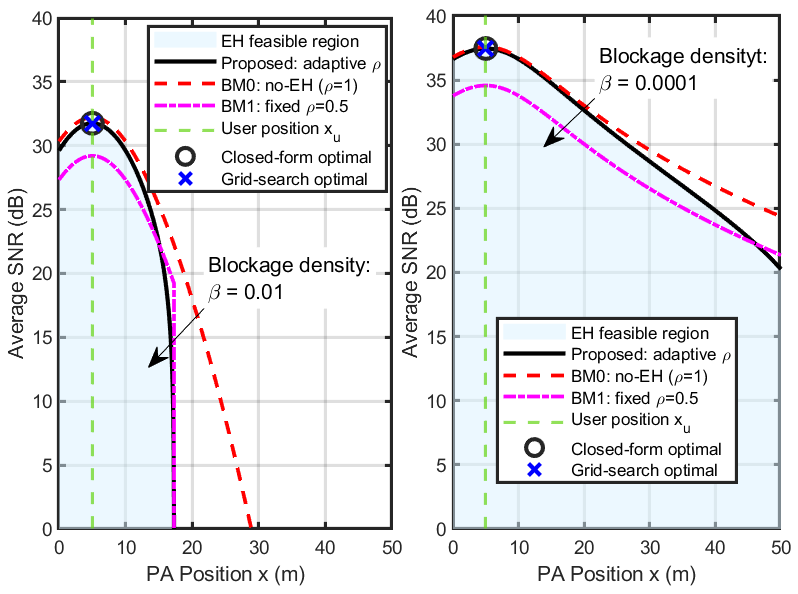}
      \label{fig_1}
  }
  \hfill
  \subfigure[]{
      \includegraphics[width=0.23\textwidth]{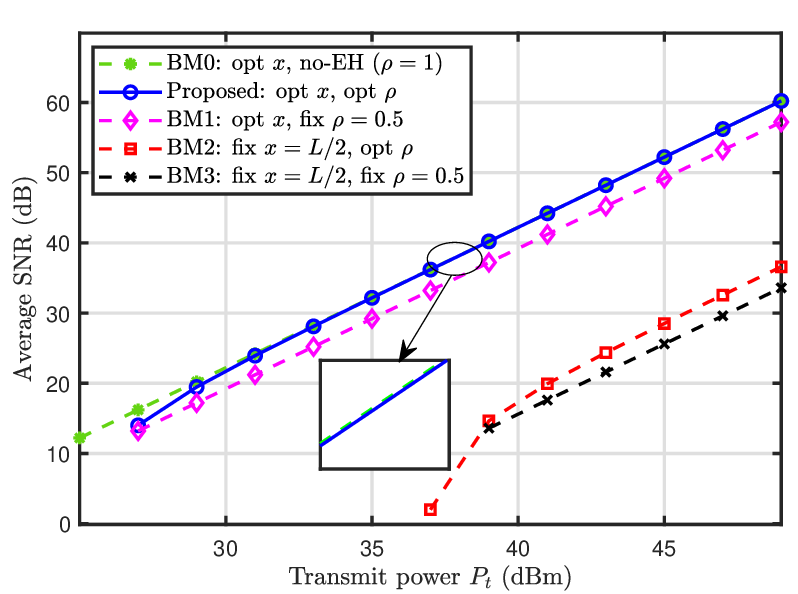}
      \label{fig_2}
  }
  \hfill
  \subfigure[]{
      \includegraphics[width=0.23\textwidth]{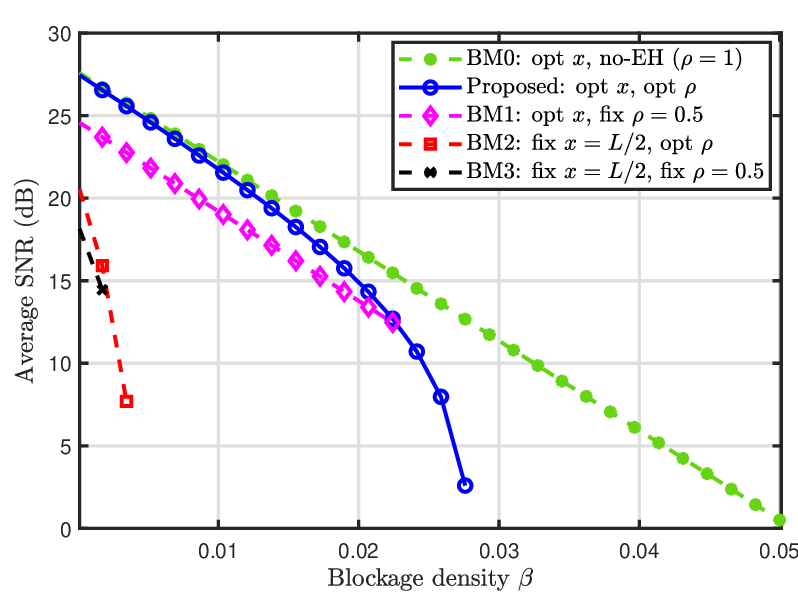}
      \label{fig_3}
  }
  \hfill
  \subfigure[]{
      \includegraphics[width=0.23\textwidth]{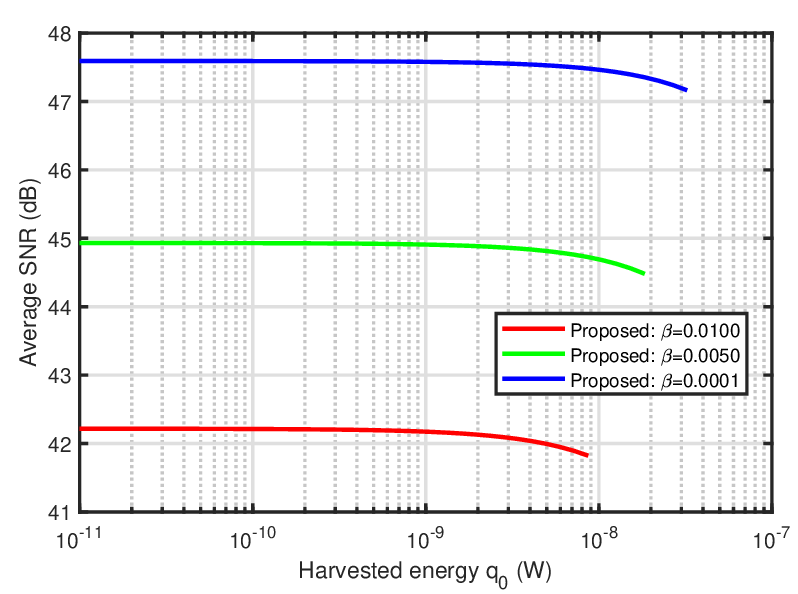} 
      \label{fig_4}
  }
  \caption{
    (a) Average SNR vs. PA position $x$ under different blockage densities $\beta$: the proposed scheme, BM0 (no-EH), and BM1 (fixed $\rho=0.5$).
    (b) Average SNR vs. $P_{\rm t}$ under $\beta=0.01$.
    (c) Average SNR vs. blockage density $\beta$ with $P_{\rm t} = 30$ dBm. 
    (d) Average SNR vs. EH requirement $q_0$.}
\end{figure*}

\section{Simulation Results}

In this section, we present numerical results to validate the theoretical analysis developed in the preceding sections. Unless otherwise specified, the default simulation parameters are set as follows, based on \cite{Ding2025LoS,xu2025blockage,xu2025pinching}. The system operates at a carrier frequency of $f_c = 28$ GHz. The transmit power is set to $P_{\rm t} = 40$ dBm, the noise variance to $\sigma^2 = 10^{-11}$ W, and the energy conversion efficiency to $\zeta = 0.6$. The PA moves along a horizontal segment of length $L = 50$ m at a fixed waveguide height of $z_{\rm p} = 10$ m. The user is located at $(x_{\rm u}, y_u, 0) = (5, 5, 0)$ m. The blockage intensity $\beta$ takes values in $\{10^{-2}, 10^{-3}, 10^{-4}\}$, and the minimum required harvested power $q_0$ ranges from $10^{-11}$ to $10^{-7}$~W. For comparison, the following benchmark schemes (BM) are considered: 
\begin{itemize}
    \item \textbf{Proposed}: jointly optimize $x$ and $\rho$.
    \item \textbf{BM0}: No-EH, i.e.,  Optimize $x$ with fixed $\rho = 1$;
    \item \textbf{BM1}: Optimize $x$ with fixed $\rho = 0.5$.
    \item \textbf{BM2}: Fix $x = L/2$ and optimize $\rho$.
    \item \textbf{BM3}: Uniform allocation with $x = L/2$ and $\rho = 0.5$.
\end{itemize}






Fig. \ref{fig_1} compares the average SNR versus PA position $x$ under different blockage densities $\beta$ among the proposed scheme, BM0, and BM1. It is seen that the closed-form solution and the numerically optimized result via the Grid-section search coincide exactly, validating the global optimality of the analysis. As $\beta$ decreases, the LoS blockage effect weakens, leading to a larger feasible region (shaded area) and enabling the optimal PA position $x$ to move closer to $x_{\rm u}$, where channel gain is maximized. Moreover, in the feasible region, the proposed scheme closely follows the no-EH baseline (BM0), indicating minimal impact from the EH constraint. Outside this region, the SNR drops sharply due to insufficient EH. In contrast, the fixed $\rho=0.5$ baseline (BM1) performs worse, highlighting the benefits of the adaptive approach.


Fig.~\ref{fig_2} shows the average SNR versus transmit power $P_{\rm t}$. We can see that the proposed scheme closely approaches the performance of BM0 at high $P_{\rm t}$, demonstrating an effective balance between EH and ID. At low $P_{\rm t}$, the gap to BM0 is small, as EH constraints are easily satisfied. As $P_{\rm t}$ increases, the gap slightly widens due to the growing power allocation required for EH. In contrast, benchmarks with fixed $x$ or $\rho$ exhibit significant performance degradation, especially at low $P_{\rm t}$, where EH constraints severely limit feasible operation.

Fig.~\ref{fig_3} plots the average SNR versus blockage density $\beta$ for $P_t = 30\,$dBm. It is observed that the proposed scheme closely follows the performance of BM0 at low $\beta$, where channel conditions are favorable and the EH constraint is easily met. As $\beta$ increases, the gap widens due to the growing difficulty in satisfying the EH requirement, which forces a higher PS ratio for EH and reduces the signal power available for ID. In contrast, schemes with fixed $x$ or $\rho$ show poorer performance, highlighting the importance of joint optimization under dynamic blockage conditions.

Fig.~\ref{fig_4} shows the average SNR versus the EH requirement $q_0$ under varying LoS blockage densities $\beta$. It is seen that as $q_0$ increases, the average SNR decreases across all $\beta$ values. Because a higher $q_0$ imposes stricter EH constraints, forcing the system to allocate more power to EH, leaving less for ID and degrading the average SNR. Moreover, at large $q_0$, the feasible region for $(x, \rho)$ shrinks or even vanishes, leading to performance saturation or infeasibility.


\begin{figure*}[htp!]
  \centering
  \subfigure[]{
      \includegraphics[width=0.23\textwidth]{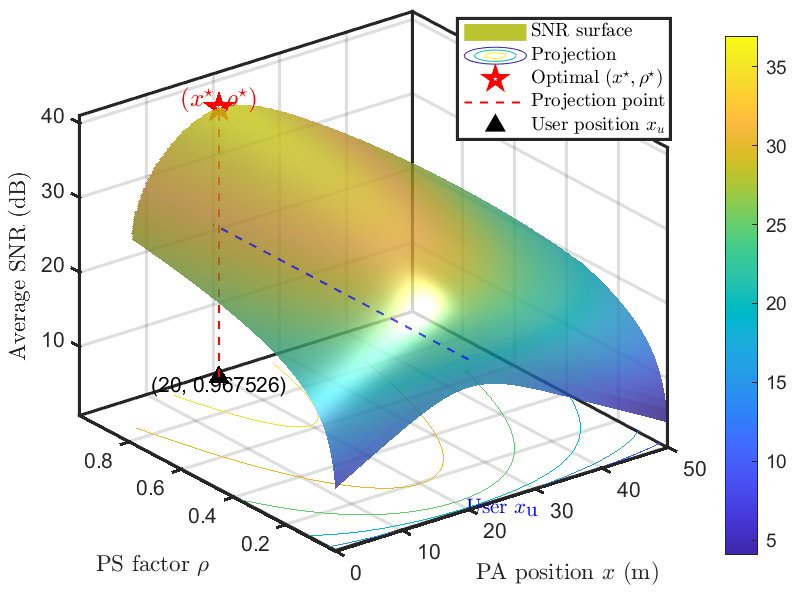}
      \label{fig_5}
  }
  \hfill
  \subfigure[]{
      \includegraphics[width=0.23\textwidth]{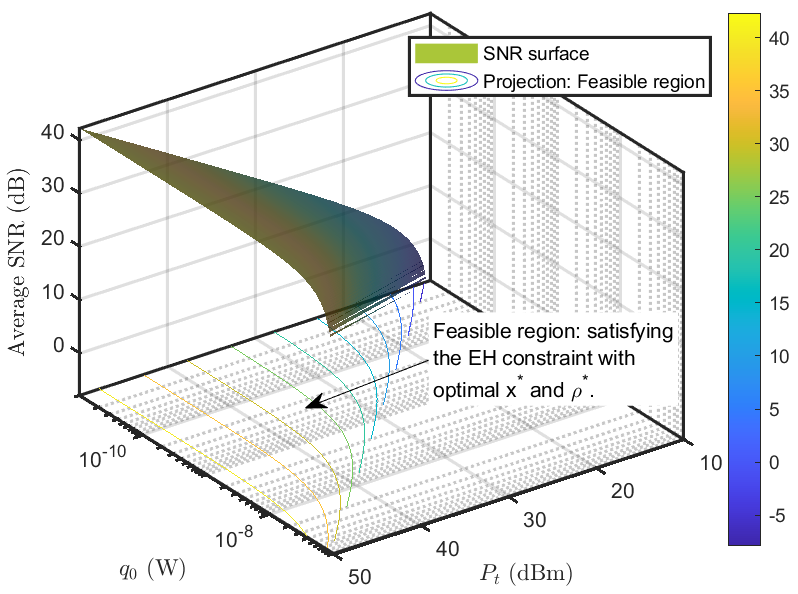}
      \label{fig_6}
  }
  \hfill
  \subfigure[]{
      \includegraphics[width=0.23\textwidth]{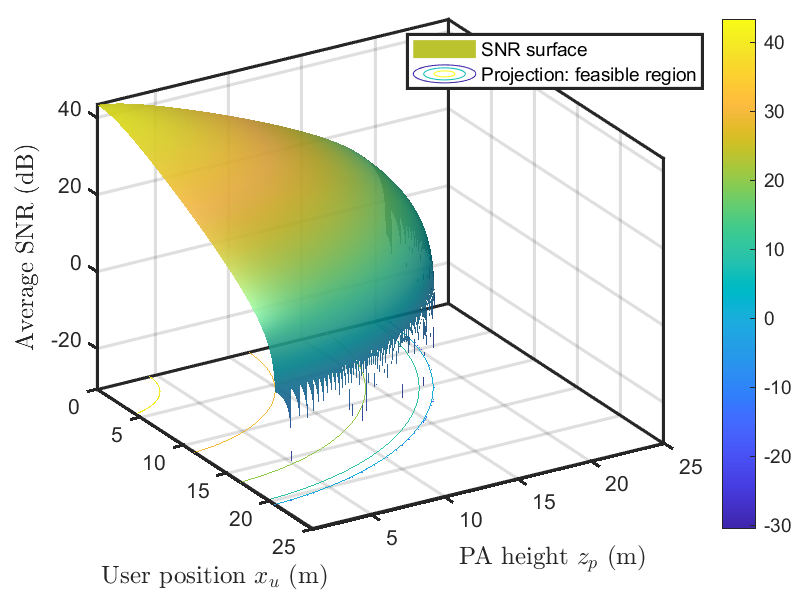}
      \label{fig_7}
  }
  \hfill
  \subfigure[]{
      \includegraphics[width=0.23\textwidth]{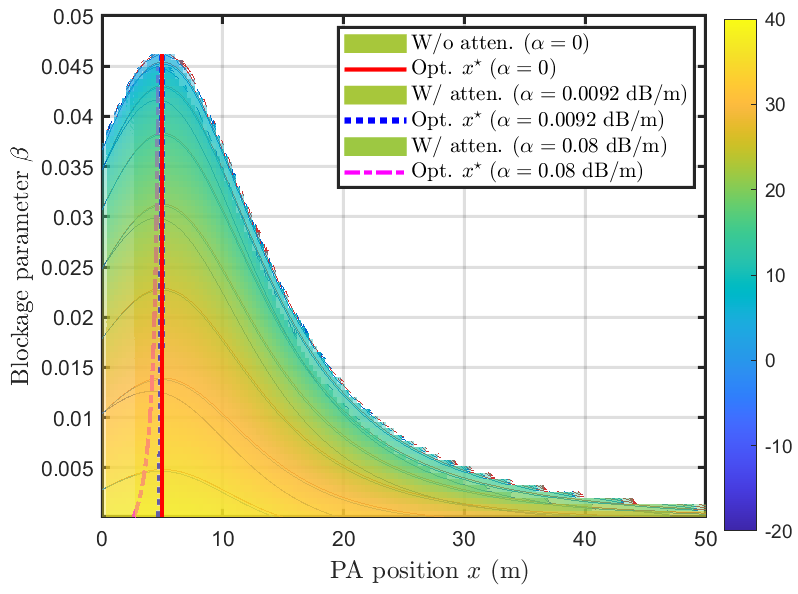} 
      \label{fig_8}
  }
  \caption{
    (a) Average SNR versus $x$ and $\rho$ under $\beta=0.001$.
    (b) Average SNR vs. $P_{\rm t}$ and EH requirement $q_0$ under $\beta=0.01$.
    (c) Average SNR vs. user position $x_{\rm u}$ and PA height $z_{\rm p}$ under $\beta=0.01$.
    (d) Average SNR versus PA position $x$ and blockage parameter $\beta$ under different waveguide attenuation $\alpha$.
  }
\end{figure*}




Fig.~\ref{fig_5} shows the average SNR versus PA position $x$ and PS ratio $\rho$ with $\beta = 0.001$. It is seen that the optimal point $(x^\star, \rho^\star)$ is close to the user position $x_{\rm u}$, consistent with our analysis. At low $\beta$, signal attenuation is weak, allowing the PA to be placed near $x_{\rm u}$ without violating the EH constraint. The relatively high $\rho^\star$ indicates more power allocated to ID while still meeting the EH requirements. Low blockage enhances received power, relaxing the EH constraint and enabling both maximum channel gain and higher SNR through increased $\rho^\star$.

Fig.~\ref{fig_6} shows the average SNR versus $P_{\rm t}$ and $q_0$ under $\beta=0.01$. It is seen that the average SNR increases with $P_{\rm t}$, as higher transmit power improves both ID and EH. Further, the feasible region shrinks at high $q_0$ and low $P_{\rm t}$, where the EH constraint cannot be satisfied. The surface exhibits a sharp drop near the feasibility boundary, reflecting the tight coupling between $q_0$ and $P_{\rm t}$, as discussed in Remark \ref{rem1}. 

Fig. \ref{fig_7} shows the average SNR versus the user position $x_{\rm u}$ and PA height $z_{\rm p}$. We see that the average SNR peaks when the user is near the feed point ($x_{\rm u} \approx 0$) and the PA is mounted at a moderate height ($z_{\rm p} \approx 5$--10\,m), decreasing as either increases. This is due to increased path loss from free-space spreading and blockage attenuation at larger distances, reducing channel gain. To satisfy the EH constraint, further limiting ID power in weak channels. The peak reflects the trade-off between proximity, propagation loss, and EH demand.

Fig. \ref{fig_8} plots the average SNR versus the PA position $x$ and blockage parameter $\beta$ under different waveguide attenuation $\alpha$. We can observe that the SNR peaks when the PA is placed near the user and decreases rapidly as the PA moves away, due to the increased path loss and waveguide transmission loss. Higher $\alpha$ leads to lower SNR and a slight shift of $x^\star$ toward the feed point, because larger $\alpha$ introduces more transmission loss and requires a shorter waveguide distance to compensate. Moreover, increasing $\beta$ degrades the SNR with little change in $x^\star$, as stronger blockage just reduces the received power but does not affect the optimal trade-off between path loss and waveguide loss. The results demonstrate that placing the PA near the user can effectively maximize SNR, and reducing $\alpha$ is critical to expanding the effective communication range.

\section{Conclusion}
In this paper, we have proposed a PA-enabled SWIPT system with a PS receiver in the probabilistic LoS blockage environment. We have formulated a joint optimization of PA position and PS ratio to maximize average SNR subject to EH and PA placement constraints. We have derived and validated a closed-form solution as the global optimum. The results can provide a foundation for energy-efficient reconfigurable antenna systems in probabilistic LoS blockage scenarios. Future work will explore multi-user scenarios, non-linear EH models, and more configurations of PA systems.

\appendices



\end{document}